\documentclass[aps,prl,reprint,superscriptaddress]{revtex4-1}

\usepackage{graphicx,amsmath,xcolor}

\def\F{{Fr\"ohlich}}

\def\oden{Oden Institute for Computational Engineering and Sciences,
The University of Texas at Austin, Austin, Texas 78712, USA}
\def\physics{Department of Physics, The University of Texas at Austin, Austin, Texas 78712, USA}
\def\macau{Institute of Applied Physics and Materials Engineering,
University of Macau, Macao SAR 999078, P. R. China}
\def\ubc{Fisika Saila, University of the Basque Country UPV/EHU, 48080 Bilbao, Basque 
Country, Spain}
\def\dipc{Donostia International Physics Center (DIPC), Paseo Manuel de Lardizabal 4, 20018 Donostia-San 
Sebasti\'{a}n, Spain}
\def\EHUq{EHU Quantum Center, University of the Basque Country UPV/EHU, Barrio Sarriena, s/n, 48940 Leioa, Biscay, Spain}

\begin{document}

\title{Unified approach to polarons and phonon-induced band structure renormalization}

\author{Jon Lafuente-Bartolome}
\author{Chao Lian}
  \affiliation{\oden}
  \affiliation{\physics}
\author{Weng Hong Sio}
  \affiliation{\macau}
\author{Idoia G. Gurtubay}
\author{Asier Eiguren}
  \affiliation{\ubc}
  \affiliation{\dipc}
  \affiliation{\EHUq}
\author{Feliciano Giustino}
  \email{fgiustino@oden.utexas.edu}
  \affiliation{\oden}
  \affiliation{\physics}

\date{\today}

\begin{abstract}
\textit{Ab initio} calculations of the phonon-induced band structure renormalization are currently based 
on the perturbative Allen-Heine theory and its many-body generalizations. These approaches are unsuitable 
to describe materials where electrons form localized polarons. Here, we develop a self-consistent, many-body
Green's function theory of band structure renormalization that incorporates localization and self-trapping. 
We show that the present approach reduces to the Allen-Heine theory in the weak-coupling limit, and to total 
energy calculations of self-trapped polarons in the strong-coupling limit. To demonstrate this methodology, we 
reproduce the path-integral results of Feynman and diagrammatic Monte Carlo calculations for the \F\ model 
at all couplings, and we calculate the zero point renormalization of the band gap of an ionic insulator 
including polaronic effects.
\end{abstract}

\maketitle

\begin{figure*}
    \includegraphics[width=\textwidth]{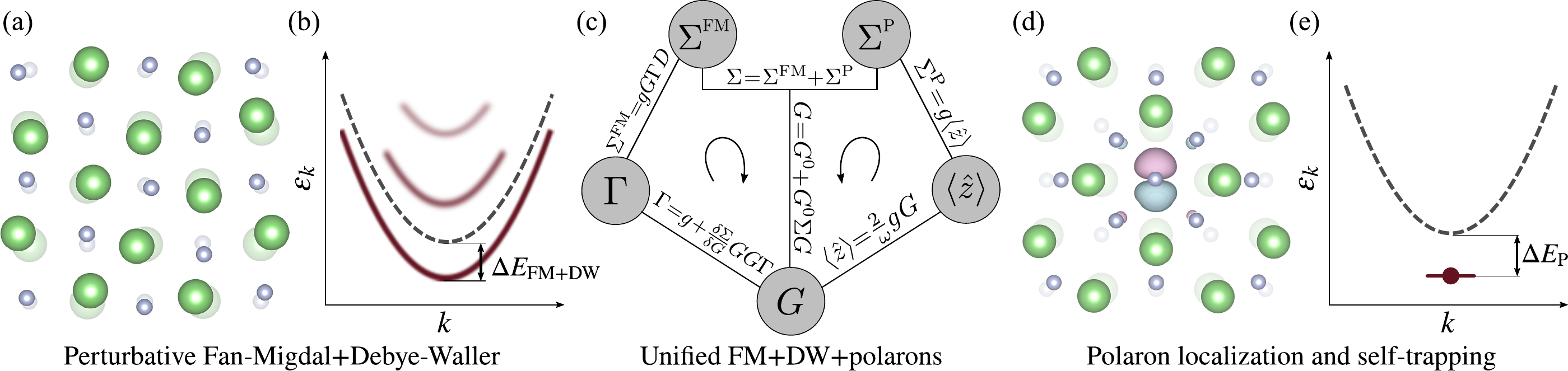}
    \caption{(a) Schematic illustration of the ground state 
    of the $N$-electron system, with atoms vibrating around the equilibrium sites of the periodic crystal. 
    (b) Schematic of phonon-induced band structure renormalization, as obtained by using the Fan-Migdal and Debye-Waller self-energies. 
    The dashed line is the non-interacting band, the brown lines are the renormalized band and its phonon sidebands.
    (c) Self-consistent set of equations for calculating electron-phonon renormalization of band structures
    including polaron localization effects, Eqs.~\eqref{eq:elph_ham}-\eqref{eq:LP_selfen}.  (d) Schematic 
    illustration of the ground state of the $N\!+\!1$-electron system, where the excess electron forms a localized 
    polaron. (e) In the scenario illustrated in (d), the energy of the conduction band bottom is lowered by the 
    formation energy of the polaron.
    \label{fig1}}
\end{figure*}

The past decade has seen much progress in first-principles calculations of phonon-induced renormalization
of band structures, including temperature dependence and quantum zero-point effects \cite{Grimvall1981,GiustinoRMP2017}.
For example, since the initial \textit{ab initio} implementations \cite{MariniPRL2008,GiustinoPRL2010}
of the Allen and Heine (AH) theory \cite{AllenHeineJPC1976}, several improvements have been made including
calculations of complete band structures of semiconductors 
\cite{PoncePRB2014,PonceJCP2015,CarusoPRB2019,BrownPRB2020} 
and non-adiabatic 
effects \cite{NeryPRB2016,MiglioNPJ2020}. On a related front, \textit{ab initio} many-body Green's function approaches have 
been used to calculate \cite{EigurenPRL2003,ParkPRL2007,GiustinoNature2008,EigurenPRB2009,
VerdiNCOM2017,RileyNCOM2018,NeryPRB2018,
GoiricelayaCMP2019,ZhouPRR2019,AntoniusPRR2020,LiPRL2021} band structure kinks and satellites observed in angle resolved 
photoelectron spectra \cite{DamascelliRMP2003,LanzaraNAT2001,
MoserPRL2013,ChenNCOM2015, WangNMAT2016,CancellieriNCOM2016,
KangNMAT2018}, cf.\ Fig.~\ref{fig1}(a),(b).  
One important limitation of these methods is that they do not 
consider the possibility of electron localization into a polaron. 

\raggedbottom

A polaron forms when an excess electron induces a distortion of the crystal lattice, which in turn acts as a
potential well and promotes electron localization \cite{Alexandrov2008,Devreese2020arXiv,FranchiniNREVMAT2021}. 
Calculations of polarons are usually performed by adding or removing an electron from a large supercell using 
density-functional theory (DFT) \cite{FranchiniPRL2009,DeskinsPRB2007,LanyPRB2009,VarleyPRB2012, SadighPRB2015, 
KokottNJP2018, LeePRM2021}, cf.\ Fig.~\ref{fig1}(d),(e).  To overcome the DFT self-interaction error and the 
computational complexity of large supercell calculations, this direct approach has recently been reformulated 
as a nonlinear eigenvalue problem within density-functional perturbation theory (DFPT) \cite{SioPRL2019,SioPRB2019}.  
These ``polaronic'' methods carry two limitations: ions are described using the adiabatic Born-Oppenheimer 
approximation, and quantum nuclear effects are neglected.

The relation between AH-based approaches, which include many-body effects but do not consider electron 
localization, and polaronic approaches, which capture localization effects but do not include non-adiabaticity 
and quantum fluctuations, remains unclear. In particular, it is unclear whether these methods describe the same 
physics, so that they can be used interchangeably, or else they capture separate phenomena. Furthermore, it
is unclear whether one approach is to be preferred over another for specific classes of materials.

Here, we address these questions by developing a self-consistent many-body Green's function theory of phonon-induced 
band structure renormalization which includes non-adiabatic effects and localization on the same footing. We show 
that the present theory reduces to AH-based approaches for materials that host large polarons, and to the 
\textit{ab initio} polaron equations of Ref.~\onlinecite{SioPRB2019} for materials with small polarons. To illustrate 
the broad applicability of this method, we calculate the energy of the \F\ polaron, and we obtain very good 
agreement with the path integral results of Feynman \cite{FeynamnPR1955} and with diagrammatic Monte Carlo
calculations \cite{ProkofevPRL1998,MischenkoPRB2000}. 
As a first \textit{ab initio} calculation using this method, we obtain the phonon-induced 
band gap renormalization of LiF, and we show that polaron localization effects dominate over the standard 
Fan-Migdal and Debye-Waller self-energies \cite{GiustinoRMP2017}.

\raggedbottom

The 
effective
Hamiltonian describing a coupled electron-phonon system is given by \cite{AccompanyingPaper,GiustinoRMP2017}:
\begin{multline} \label{eq:elph_ham}
   \hat{H}
   = \sum_{n\textbf{k}} \varepsilon_{n\mathbf{k}} \hat{c}_{n\mathbf{k}}^\dagger \hat{c}_{n\mathbf{k}} +
   \sum_{\mathbf{q}\nu} \hbar\omega_{\mathbf{q}\nu} (\hat{a}_{\mathbf{q}\nu}^\dagger 
\hat{a}_{\mathbf{q}\nu}+1/2) \\
   ~~~+ N_p^{-\frac{1}{2}} \sum_{\substack{\mathbf{k},\mathbf{q} \\ m n \nu }} 
g_{mn\nu}(\mathbf{k},\mathbf{q}) \,
   \hat{c}_{m\mathbf{k}+\mathbf{q}}^\dagger \hat{c}_{n\mathbf{k}} 
(\hat{a}_{\mathbf{q}\nu}+\hat{a}_{-\mathbf{q}\nu}^\dagger) ~,
\end{multline}
where $\varepsilon_{n\mathbf{k}}$ represents the single-particle eigenvalue of an electron in the band $n$ with 
crystal momentum $\mathbf{k}$, $\omega_{\mathbf{q}\nu}$ is the frequency of a phonon in the branch $\nu$ with 
crystal momentum $\mathbf{q}$, and $\hat{c}^\dagger_{n\mathbf{k}}/\hat{c}_{n\mathbf{k}}$ and 
$\hat{a}^\dagger_{\mathbf{q}\nu}/\hat{a}_{\mathbf{q}\nu}$ are the associated fermionic and bosonic 
creation/annihilation operators, respectively; $g_{mn\nu}(\mathbf{k},\mathbf{q})$ denotes the electron-phonon 
coupling matrix element between the electrons $n\mathbf{k}$ and $m\mathbf{k+q}$ via the phonon $\mathbf{q}\nu$, 
and $N_p$~is the number of unit cells in the periodic Born-von K\'arm\'an supercell.
The limitations of 
the effective Hamiltonian in Eq.~(\ref{eq:elph_ham}) are discussed in 
the companion manuscript \cite{AccompanyingPaper}

To investigate the ground state of the Hamiltonian in Eq.~\eqref{eq:elph_ham} in the presence of an excess
electron or hole, we focus on the electron Green's function. We consider a periodic crystal with $N$ electrons, 
and we define the electron Green's function as the expectation value of the field operators over the ground 
state of the $N\!+\!1$-particle system: $G_{12} = -(i/\hbar) \langle N\!+\!1| \hat{T} \, \hat{c}_{1} \, 
\hat{c}^\dagger_{2} | N\!+\!1 \rangle$. In this definition we use the compact notation $1 = \{n_1,\mathbf{k}_1, 
t_1\}$ and $2 = \{n_2,\mathbf{k}_2, t_2\}$, $t$ is the time, and $\hat{T}$ is the time-ordering operator. 
Our present definition of Green's function differs from the conventional definition \cite{HedinLundqvistSSP1969} 
where the expectation value is over the the ground state $|N \rangle$; this choice is essential to capture 
localization effects. Using Schwinger's functional derivative technique \cite{KatoPTP1960,HedinLundqvistSSP1969,GiustinoRMP2017}, 
in the companion manuscript \cite{AccompanyingPaper} we derive the following Dyson equation:
\begin{equation} \label{eq:Dyson}
    G_{12} = G^0_{12} + G^0_{13} \left(\Sigma^{\mathrm{FM}}_{34} + \Sigma^{\mathrm{P}}_{34}\right) G_{42},
\end{equation}
where summation over repeated numbered indices is implied
throughout the manuscript. 
In this expression, $G^0$ is the Green's function in the
absence of electron-phonon interactions, $\Sigma^{\mathrm{FM}}$ is the Fan-Migdal self-energy \cite{FanPR1951,
MigdalJETP1958,EngelsbergSchriefferPR1963,GiustinoRMP2017}, and $\Sigma^{\mathrm{P}}$ is a new contribution 
which we call ``polaronic'' self-energy.

The Fan-Migdal self-energy is given by:
\begin{equation} \label{eq:FM_selfen}
    \Sigma^{\mathrm{FM}}_{12}
    =  i\, g_{314} \, G_{3(1),5} \, \Gamma_{526} \, D_{6,4(1)}~,
\end{equation}
where the electron-phonon matrix elements is written compactly as ${g_{123} = N_p^{-1/2} g_{n_2n_1\nu_3}
(\mathbf{k}_1,\mathbf{q}_3)}\, \delta_{\mathbf{k}_2,\mathbf{k}_1+\mathbf{q}_3}$, 
the notation $G_{3(1),5}$ 
stands for $G_{n_3\mathbf{k}_3,n_5 \mathbf{k}_5}(t_1,t_5)$, and there is no summation over bracketed indices.
In Eq.~\eqref{eq:FM_selfen}, $D$ is the phonon Green's function and $\Gamma$ is the electron-phonon vertex; 
explicit expressions for these quantities are provided in Ref.~\onlinecite{AccompanyingPaper}.

The polaronic self-energy $\Sigma^{\mathrm{P}}$ appearing in Eq.~\eqref{eq:Dyson} is given by:
\begin{align} \label{eq:LP_selfen}
    \Sigma^{\mathrm{P}}_{12}
    &=
    \delta(t_1\!-\!t_2)
    \, g_{213}
    \, \frac{\langle \hat{z}_3 \rangle}{l_3}~.
\end{align}
In this equation, 
$l_3=l_{\mathbf{q}\nu}$ is a short for the zero-point displacement amplitude, and
the term 
$\langle \hat{z}_3 \rangle=\langle \hat{z}_{\mathbf{q}\nu} \rangle$
represents the expectation value of the 
normal vibrational coordinates $\hat{z}_{\mathbf{q}\nu}$ over the ground state of the $N+1$-particle system, 
$\langle N\!+\!1| \hat{z}_{\mathbf{q}\nu} | N\!+\!1\rangle$
, which is directly related to the atomic displacements in the polaronic configuration \cite{AccompanyingPaper}. 
This expectation value depends in turn on the 
many-body electron density via the equal-time Green's function,
$\langle \hat{z}_3 \rangle=-il_3(2/\omega_3)g^{*}_{453}G_{5(1),4(1^+)}$ \cite{AccompanyingPaper}.  
{$\Sigma^\mathrm{P}$ is} nonzero whenever 
the atoms of the $N\!+\!1$-electron ground state are displaced from the equilibrium sites of the $N$-electron 
ground state, hence it describes polaron localization effects.

Equations~\eqref{eq:elph_ham}-\eqref{eq:LP_selfen} define a self-consistent formulation of the electron-phonon
renormalization of energy bands which includes the effects of polaron formation. The relation between these
equations is schematically illustrated in Fig.~\ref{fig1}(c). 

\begin{figure}
    \includegraphics[width=1.0\columnwidth]{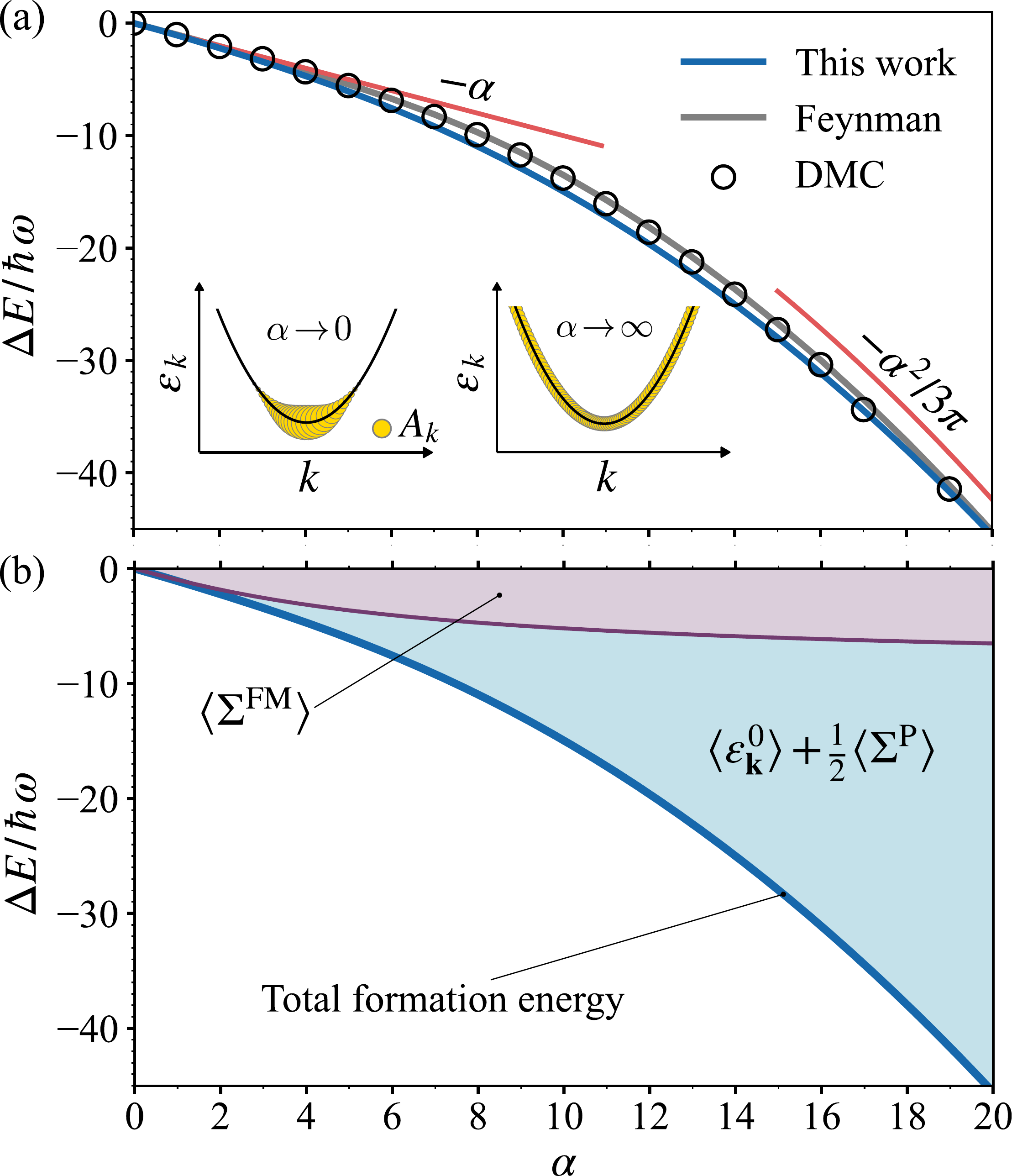}
    \caption{(a) Ground-state energy of the \F\ polaron, $\Delta E/\hbar\omega$, as a function of the coupling 
	strength $\alpha$: present calculation (blue line), Feynman's path integral results \cite{FeynamnPR1955} (gray line),
    and diagrammatic Monte Carlo (DMC) data taken from Ref.~\cite{HahnPRB2018} (black circles). 
    Red lines indicate the asymptotic expansions at weak 
    and strong coupling, respectively. The quasiparticle amplitudes $|A_{k}|^2$ in these limits are shown
    in the inset, superimposed to the free electron band.  (b) \mbox{Breakdown} of the ground-state energy of the 
    \F\ polaron into its self-energy contributions.
    \label{fig2}}
\end{figure}

\begin{figure*}
    \includegraphics[width=1.0\linewidth]{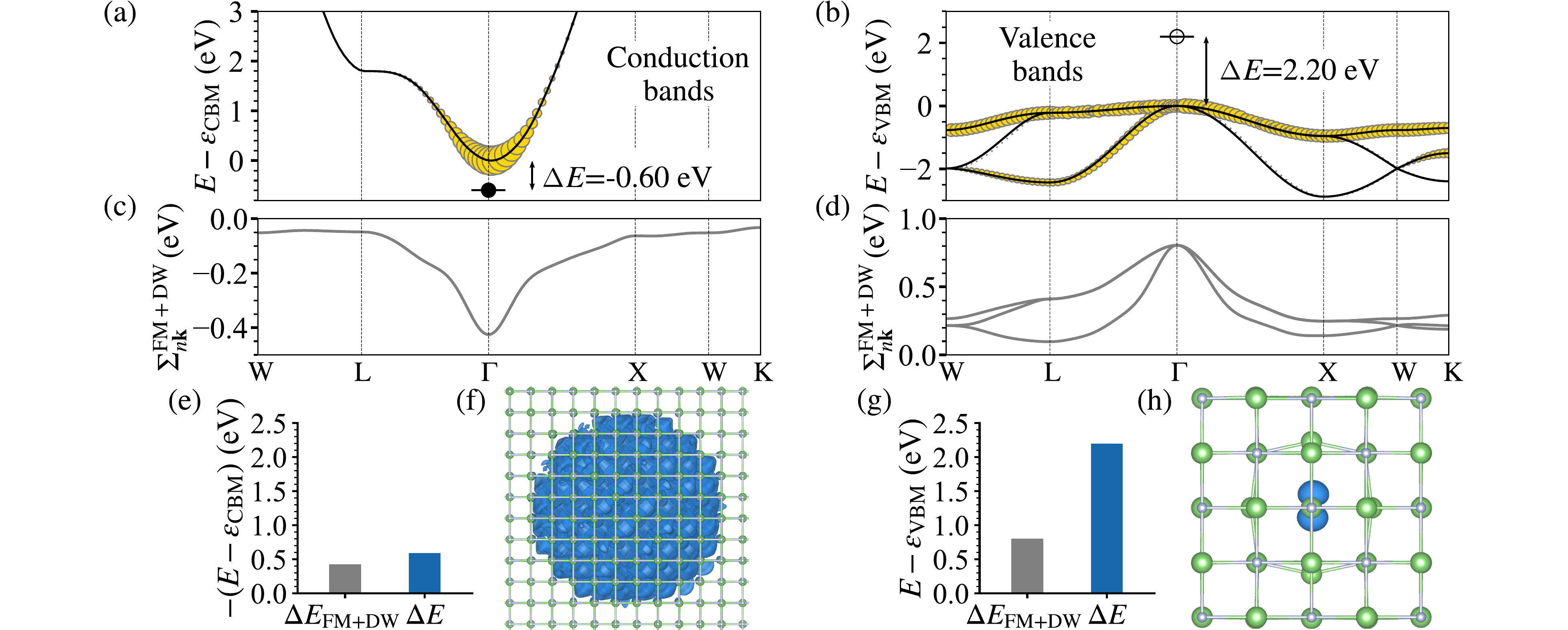}
    \caption{(a), (b) Conduction and valence bands of 
    LiF, with energies referred to the conduction band minimum (CBM) and to the valence band maximum (VBM), respectively. 
    The yellow disks indicate the square moduli of the quasiparticle amplitudes of the Dyson orbitals, as obtained 
    by solving Eqs.~(5), (8), and (9).
    In each panel we indicate the quasiparticle 
    renormalization including polaronic effects, $\Delta E$.
    (c), (d) Expectation values 
    of $\Sigma^{\rm FM}+\Sigma^{\rm DW}$ along the bands.  (e), (g) Calculated renormalization of the conduction 
    and valence band extrema, respectively, using two methods: the standard perturbative approach which does not 
    include polaron localization (gray), and the present approach including localization effects (blue).  
    (f), (h) Calculated wavefunctions of the electron and hole polarons in LiF, respectively.
    \label{fig3}}
\end{figure*}

Since our formalism starts from Eq.~\eqref{eq:elph_ham}, which does not include terms of second order
in the atomic displacements, our self-energy $\Sigma$ does not contain the standard Debye-Waller contribution. 
This term must be added separately, as described in the companion manuscript~\cite{AccompanyingPaper}.

The above formalism can be turned into a practical computational method by expressing 
the Fourier transform of the Green's function into frequency domain $\omega$ 
via Dyson orbitals $f_s(\mathbf{r})$ using the Lehmann representation, 
$G(\omega) = \sum_s f_s f_s^* /[\hbar\omega - \varepsilon_s - i\eta \mathrm{sgn}(\mu-\varepsilon_s) ]$,
where $\mu$ is the chemical potential, and $\eta\rightarrow 0^+$.
The Dyson orbitals of the occupied manifold are given by $f_{s}(\mathbf{r}) = \langle \, N,s \,| \, \hat{\psi}(\mathbf{r}) 
\, | \,N\!+\!1 \, \rangle$ \cite{HedinLundqvistSSP1969,AccompanyingPaper}, 
where $|N,s\rangle$ denotes the $s$-th excited state of the $N$-electron system, 
$\hat{\psi}(\mathbf{r})$ is the electron field operator,
and $\varepsilon_s=E_{N+1}-E_{N,s}$.
In Ref.~\onlinecite{AccompanyingPaper} we show that we can identify the Dyson orbital 
for the lowest-energy excitation of the $N\!+\!1$-particle system with the electronic component of the polaron 
wavefunction.  
Following 
the strategy of Ref.~\onlinecite{SioPRB2019}, we expand the orbitals in the basis of single-particle Bloch 
wavefunctions $\psi_{n\mathbf{k}}$, ${f_{s} = N_{p}^{-1/2} \sum_{n\mathbf{k}} A_{n\mathbf{k}}^{s} \, \psi_{n\mathbf{k}}}$.
This representation allows us to recast Eqs.~(\ref{eq:Dyson})-(\ref{eq:LP_selfen}) into a nonlinear eigenvalue 
problem for the quasiparticle amplitudes $A_{n\mathbf{k}}^{s}$ and the 
electron addition/removal
energies $\varepsilon_s$:
\begin{equation} \label{eq:qp_eq}
    \left[ \varepsilon_1 \delta_{12}
    + \Sigma^{\mathrm{FM}}_{12}(\varepsilon_s/\hbar)
    + \Sigma^{\mathrm{P}}_{12} \right]
    A_2^{s}
    = \varepsilon_s A_1^s ~,
\end{equation}
where
\begin{multline}
    \Sigma^{\mathrm{FM}}_{12}(\omega) \!=\!
    \pm g_{143}^* \, g_{253} 
    \sum_s
    \, \frac{A_4^{s} A_5^{s,*} }{N_p}
    \frac{\theta\left[\pm(\varepsilon_{s}-\mu)\right]}{
    \pm \hbar\omega \mp \varepsilon_{s}\! - \! \hbar \omega_{3} \! + \! i\eta} 
    , \label{eq:FM_As}
\end{multline}
\begin{equation}    
    \Sigma^{\mathrm{P}}_{12} =  
    -\frac{2 \, g_{213} \, g_{543}^*}{\hbar\omega_3} 
    \sum_{s}^{\varepsilon_s < \mu} 
    \frac{A_4^{s} \, A_5^{s,*}}{N_p} ~. \label{eq:LP_As}
\end{equation}
In Eq.~\eqref{eq:FM_As}, there is a sum over the $\pm$ terms
and $\theta$ is the Heaviside step function.
The same expressions are given without using
compact notation in Eqs. (41) and (45) of the companion manuscript \cite{AccompanyingPaper}.
To reach Eq.~\eqref{eq:FM_As} we approximated the vertex $\Gamma$ by the standard electron-phonon matrix element 
$g$, and we replaced the interacting phonon Green's function $D$ by its 
adiabatic
counterpart, as obtained 
e.g. from DFPT calculations. 

Equations~(\ref{eq:qp_eq})-(\ref{eq:FM_As}) are still too complex for \textit{ab initio} calculations. To proceed 
further, we assume that the added electron in the $(N\!+\!1)$-electron system has a negligible effect on the 
valence manifold of the $N$-electron system. 
The validity of this assumption is assessed in Ref.~\cite{AccompanyingPaper}.
With this choice, the $N$ occupied Dyson orbitals can be replaced by Bloch wave functions,
and their contribution to $\Sigma^\mathrm{P}$ vanishes, while the Dyson orbital of the excess electron is to be 
determined by solving the equations self-consistently. 
After this simplification,
and replacing the Green's function by its non-interacting counterpart in Eq.~\eqref{eq:FM_As},
Eqs.~(\ref{eq:FM_As}) and (\ref{eq:LP_As}) become: 
\begin{multline}
        \Sigma^{\mathrm{FM}}_{n\mathbf{k},n'\mathbf{k'}}(\omega)
        = 
        \pm\frac{\delta_{n\mathbf{k},n'\mathbf{k'}}}{N_{p}} \sum_{m \mathbf{q} \nu}
        |g_{mn\nu}(\mathbf{k},\mathbf{q})|^2 \\
        \times
        \frac{\theta\left[\pm(\varepsilon_{m \mathbf{k+q}}-\mu)\right]}{\pm \hbar\omega \mp \varepsilon_{m \mathbf{k+q}} - \hbar\omega_{\mathbf{q}\nu} + i\eta} ~,  \label{eq:FM_simple}
\end{multline}
\begin{multline}
    \Sigma^{\mathrm{P}}_{n\mathbf{k},n'\mathbf{k'}} 
    = - \frac{2}{N_{p}^{2}}
    \sum_{\substack{mm' \nu {\mathbf k}'' }}
    \!\!A_{m'\mathbf{k}''+\mathbf{k-k'}} \, A_{m\mathbf{k}''}^{*} 
      \\
    \times 
    \, \frac{g_{m'm\nu}^{*}(\mathbf{k}'',\mathbf{k-k'}) \, g_{nn'\nu}(\mathbf{k'},\mathbf{k-k'}) }{\hbar\omega_{\mathbf{k-k'}\nu}}~.
    \label{eq:LP_simple}
\end{multline}
These equations can be solved by using electron band structures, phonon dispersions, and
electron-phonon matrix elements from DFT and DFPT, as we show below. Once obtained the Dyson orbital
and quasiparticle eigenvalue by solving Eqs.~\eqref{eq:qp_eq}, \eqref{eq:FM_simple}-\eqref{eq:LP_simple}, we determine
the ground-state energy of the $(N\!+\!1)$-electron system using a generalized Galitskii-Migdal formula 
\cite{GalitskiiMigdalJETP1958} that we derived in Ref.~\onlinecite{AccompanyingPaper} for the coupled 
electron-phonon Hamiltonian in Eq.~(\ref{eq:elph_ham}).

The Fan-Migdal self-energy in Eq.~\eqref{eq:FM_simple} is diagonal in the electron wavevector, therefore this term does not contribute
to electron localization. Thus, the shape of the polaron quasiparticle is determined by the polaronic
term, and in the lowest-order approximation we can evaluate Eqs.~\eqref{eq:FM_simple}-\eqref{eq:LP_simple}
using a simplified procedure where we first solve for the polaron wavefunction with Eq.~\eqref{eq:LP_simple},
and then we include $\Sigma^{\rm FM}$ using perturbation theory. In Ref.~\onlinecite{AccompanyingPaper} 
we show that this procedure leads to the following expression for the 
polaronic total energy renormalization
of the system with an excess electron:
\begin{align} \label{eq:toten}
    \Delta E =
        & N_p^{-1} \! \sum_{n\mathbf{k}} 
        |A_{n\mathbf{k}}|^{2} \!
        \left[ \varepsilon_{n\mathbf{k}}^{0}
        - \varepsilon_{\mathrm{CBM}}^{0}
        \!+ \Sigma^{\mathrm{FM}}_{{n\mathbf{k}},{n\mathbf{k}}}(\omega\!=\!\varepsilon_{\mathrm{CBM}}^{0}/\hbar) \right] \nonumber \\
        &+ \frac{1}{2} \, N_p^{-1} \!\!\!\! \sum_{n\mathbf{k},n'\mathbf{k}'} \!\!\!
        A_{n\mathbf{k}}^{*}
        \, \Sigma^{\mathrm{P}}_{n\mathbf{k},n'\mathbf{k}'}\, 
        A_{n'\mathbf{k}'} ~,
\end{align}
where $\varepsilon_{\mathrm{CBM}}^{0}$ represents the energy of the conduction band minimum 
of the periodic, undistorted lattice.
This expression has an appealing physical interpretation. 
The first term on the right-hand side
is the weighted average of the
conduction band energy and the Fan-Migdal self-energy, taken over the polaron wavefunction coefficients
in reciprocal space. The last term is the stabilization energy of the electron wavefunction
resulting from the lattice distortion in the polaronic ground state. 
Therefore the total energy renormalization is
a combination of both AH-type and polaronic contributions, with their relative importance 
being dictated by the spatial extent of the wavefunction. To illustrate this point, we apply the
present methodology to the \F\ model \cite{FrohlichADP1954,Devreese2020arXiv,Alexandrov2010,Emin2012}.

The \F\ model is a standard benchmark for testing theories of coupled electrons and phonons \cite{Devreese2020arXiv}. 
It describes a free electron coupled to a dispersionless longitudinal optical phonon, with the 
coupling strength controlled by a dimensionless parameter, the \F\ coupling constant $\alpha$. AH-based 
approaches are successful in describing the weak-coupling regime ($\alpha \ll 1$) of this model, while 
polaronic approaches such as the Landau-Pekar theory \cite{LandauPZS1933,PekarZETF1946} are successful at 
strong coupling ($\alpha \gg 10$) \cite{Mahan1993}, cf.\ Fig.~\ref{fig2}(a).  State-of-the-art numerical 
results for the ground-state energy of the \F\ polaron come from diagrammatic Monte Carlo methods \cite{ProkofevPRL1998,MischenkoPRB2000,
HahnPRB2018}. As of today, the only theory that matches diagrammatic Monte Carlo results at all coupling strengths is the 
variational path integral approach by Feynman \cite{FeynamnPR1955,SchultzPR1959}. 

Figure~\ref{fig2}(a) shows the energy of the \F\ polaron as a function of $\alpha$, as calculated from 
Eq.~\eqref{eq:toten}.  The agreement between our present approach and both Feynman's solution and diagrammatic 
Monte Carlo data is very good at all couplings. In particular, our method correctly captures the expected 
linear dependence of the energy on $\alpha$ at weak coupling, 
$\Delta E = -\alpha \hbar \omega$, 
and its quadratic dependence at strong coupling, 
$\Delta E=-\alpha^2 \hbar\omega/3\pi$ \cite{Devreese2020arXiv}, cf.\ Fig.~\ref{fig2}(a).  
These limits can be rationalized by examining the 
relative contributions to the total energy shown in Fig.~\ref{fig2}(b).  At small $\alpha$, the quasiparticle 
amplitudes concentrate near the conduction band bottom (CBM), thus leading to large polarons in real space 
[cf. inset of Fig.~\ref{fig2}(a)].  In this limit, the expectation value of $\Sigma^{\mathrm{P}}$ 
tends to vanish, and the Fan-Migdal self-energy tends to $\Sigma^{\mathrm{FM}} = -\alpha \hbar\omega$. Conversely, 
at large $\alpha$ the quasiparticle amplitudes spread across the entire reciprocal space, leading to electron 
localization into a small polaron. In this limit, $\Sigma^{\mathrm{P}} = -\alpha^2 \hbar\omega /3\pi$ dominates. 

To illustrate the use of the present method for real materials, we calculate the zero-point renormalization
of rocksalt LiF, a prototypical ionic insulator. This system hosts both large electron polarons and small
hole polarons \cite{SioPRB2019}, therefore it is particularly suited to analyze the relative magnitude of
the various self-energies in the valence and conduction bands. All calculations are based on {\sc Quantum ESPRESSO} 
\cite{QE2017}, wannier90 \cite{Wannier902020}, and EPW \cite{EPW2016}, and the computational setup is described in the companion 
manuscript \cite{AccompanyingPaper}. 
We initialize the self-consistent solution of the polaron equations with a Gaussian wavepacket. 
This step is needed
to break translational symmetry, as discussed in Ref.~\cite{AccompanyingPaper}.
We verified that different initializations lead to equivalent self-consistent polaron solutions in all cases \cite{AccompanyingPaper}.

Figure~\ref{fig3} summarizes our results. In panels (a) and (b) we show the renormalization of the 
conduction band minimum and of the valence band maximum with respect to the DFT band edges, respectively. 
The quasiparticle amplitudes are represented by the solid yellow circles superimposed to the bands, with 
the radius being proportional to the square modulus $|A_{n\mathbf{k}}|^2$. In panels (c) and (d) we show
how the AH band shift varies along the conduction and valence bands, respectively. In these calculations
we evaluate the correction by including both the Fan-Migdal and the Debye-Waller self-energies, 
$\Sigma^{\mathrm{FM}}+\Sigma^{\mathrm{DW}}$ \cite{AccompanyingPaper}, to be consistent with previous 
work \cite{NeryPRB2018}.  
In both cases we see that this correction is largest at the zone center, 
and decreases towards the edges of the Brillouin zone. The localization of the polaron wavefunction softens 
this correction by averaging it over a range of wavevectors, according to the quasiparticle amplitudes 
shown in \mbox{(a) and (b)}.

In panels (e) and (g) of Fig.~\ref{fig3} we compare standard calculations of band renormalization 
using the Fan-Migdal and Debye-Waller self-energies (FM and DW) with our present approach.
In the conduction band, 
the FM and DW corrections (0.43~eV) 
are seen to yield a similar result as the total polaronic renormalization (0.60~eV). 
This finding is consistent with the observation that an excess electron in LiF forms a 
large electron polaron extended over more than ten unit cells, as shown in panel (d). 
In this scenario, 
the electron wavefunction is so delocalized that AH-based approaches provide a good description of the energy renormalization.  
Conversely, in the valence bands the polaronic renormalization (2.20~eV) 
is much larger than the FM and DW corrections (0.80~eV). 
This finding is consistent with the fact that 
an excess hole in LiF forms a small polaron, as shown in panel (h). 

By combining the above zero-point corrections for the valence and conduction band edges, we obtain
a quantum zero-point 
quasiparticle
band gap renormalization of $-2.8$~eV. This value is considerably larger than what 
one obtains by using the Fan-Migdal and Debye-Waller self-energies at the band edges, 
$-1.2$~eV.
This difference suggests that AH-based approaches may not be as reliable as previously thought in 
calculations of band gap renormalization, because they do not take into account localization
effects. We emphasize that this conclusion holds for systems that host spatially-localized
polaronic states, such as e.g. ionic compounds and oxides. 
Standard semiconductors, such as for example silicon and diamond 
\cite{MariniPRL2008,GiustinoPRL2010,PoncePRB2014,PonceJCP2015,MiglioNPJ2020}, 
do not host localized polarons, 
therefore in such cases AH-based approaches remain the current state-of-the-art.

In summary, we developed a self-consistent many-body theory of electron-phonon couplings that unifies
calculations of phonon-induced energy band renormalization and polaron localization. We found that
the lowest-order approximation to our theory matches Feynman's results for the \F\ polaron.
This methodology
is amenable to first-principles implementations, as we have demonstrated for LiF. Future work will need to
systematically assess polaronic corrections to the band renormalization of semiconductors and insulators, 
investigate the present formalism beyond the lowest-order approximation, and extend this work to calculations 
of complete band structures, phonons sidebands in ARPES spectra, 
finite-temperature effects,
and optical band gaps including excitonic effects.
We hope 
that this study will stimulate renewed efforts to understand polarons and their properties in real materials.

\begin{acknowledgments}
This research is primarily supported by the Computational Materials Sciences Program funded by the U.S. Department of 
Energy, Office of Science, Basic Energy Sciences, under Award No. DE-SC0020129 (JLB, CL, WHS: formalism, software 
development, \textit{ab initio} calculations, manuscript preparation), and by the National Science Foundation, 
Office of Advanced Cyberinfrastructure under Grant No. 2103991 (FG: project conception and supervision, manuscript 
preparation). The authors acknowledge the Texas Advanced Computing Center (TACC) at The University of Texas 
at Austin for providing HPC resources, including the Frontera and Lonestar5 systems, that have contributed 
to the research results reported within this paper.  URL: http://www.tacc.utexas.edu.  This research used 
resources of the National Energy Research Scientific Computing Center, a DOE Office of Science User Facility 
supported by the Office of Science of the U.S.  Department of Energy under Contract No. DE-AC02-05CH11231. 
WHS was supported by the Science and Technology Development Fund of Macau SAR (under Grants No. 0102/2019/A2) and the LvLiang Cloud Computing Center of China for providing extra HPC resources, including the TianHe-2 systems.
IGG and AE acknowledge the Department of Education, Universities and Research of the Eusko Jaurlaritza and the University of the Basque Country UPV/EHU (Grant No. IT1260-19), 
the Spanish Ministry of Economy and Competitiveness MINECO (Grants No. FIS2016-75862-P and No.
PID2019-103910GB-I00), 
and the University of the Basque Country UPV/EHU (Grant No. GIU18/138) for financial support.
JLB acknowledges UPV/EHU (Grant No. PIF/UPV/16/240), MINECO (Grant No. FIS2016-75862-P) and DIPC for financial support in the initial stages of this work.
\end{acknowledgments}

\bibliography{bibliography}

\end{document}